\documentclass[aps,prl,amsmath,amssymb,superscriptaddress,floatfix,showpacs]{revtex4}
\usepackage{amssymb}
\usepackage{amsfonts}
\usepackage{graphicx}
\usepackage{hyperref}
\usepackage{bm}
\begin{document}

\title{A unified data representation theory for network visualization,\\ ordering and coarse-graining}
\author{Istv\'an A. Kov\'acs\footnote{Electronic address: kovacs.istvan@wigner.mta.hu}}
\affiliation{Wigner Research Centre, Institute for Solid State Physics and Optics,
H-1525 Budapest, P.O.Box 49, Hungary}
\affiliation{Institute of Theoretical Physics, Szeged University, H-6720 Szeged, Hungary}
\affiliation{Center for Complex Networks Research and Department of Physics, Northeastern University, 110
Forsyth Street, 111 Dana Research Center, Boston, MA 02115, USA}
\author{R\'eka Mizsei}
\affiliation{Institute of Organic Chemistry, Research Centre for Natural Sciences, Hungarian Academy of Sciences, Pusztaszeri \'ut 59-67, H-1025 Budapest, Hungary}
\author{P\'eter Csermely}
\affiliation{Department of Medical Chemistry, Semmelweis University, H-1444 Budapest, P.O.Box 266, Hungary}

\date{\today}

\begin{abstract} \normalsize 
Representation of large data sets became a key question of many scientific disciplines in the last decade. Several approaches for network visualization, data ordering and coarse-graining accomplished this goal. However, there was no underlying theoretical framework linking these problems. Here we show an elegant, information theoretic data representation approach as a unified solution of network visualization, data ordering and coarse-graining. The optimal representation is the hardest to distinguish from the original data matrix, measured by the relative entropy. The representation of network nodes as probability distributions provides an efficient visualization method and, in one dimension, an ordering of network nodes and edges. Coarse-grained representations of the input network enable both efficient data compression and hierarchical visualization to achieve high quality representations of larger data sets. Our unified data representation theory will help the analysis of extensive data sets, by revealing the large-scale structure of complex networks in a comprehensible form.
\end{abstract}


\maketitle
Complex network \cite{newman,barabasi_rev} representations are widely used in physical, biological and social systems, and are usually given by huge data matrices. Network data size grew to the extent, which is too large for direct comprehension and requires carefully chosen representations. One option to gain insight into the structure of complex systems is to order the matrix elements to reveal the concealed patterns, such as degree-correlations \cite{assor,assoc} or community structure \cite{GN,newman_com,fortunato, moduland, olhede, bickel1, bickel2}. Currently, there is a diversity of matrix ordering schemes of different backgrounds, such as graph theoretic methods \cite{king}, sparse matrix techniques \cite{sparse} and spectral decomposition algorithms \cite{west}. 

Coarse-graining or renormalization of networks \cite{rg_2005,rg_2008, rozenfeld, gfeller, amaral, hier_mod} also gained significant attention recently as an efficient tool to zoom out from the network, while reducing its size to a tolerable extent. A variety of heuristic coarse-graining techniques -- also known as multi-scale approaches -- emerged, leading to significant advances of network-related optimization problems \cite{travel,opt} and the understanding of network structure \cite{rg_2008, rozenfeld,link}. 

The most essential tool of network comprehension is a faithful visualization of the network \cite{battista}. Preceding more elaborate quantitative studies, it is capable of yielding an intuitive, direct qualitative understanding of complex systems. Although being of a primary importance, there is no general theory for network layout, leading to a multitude of graph drawing techniques. Among these, force-directed \cite{force} methods are probably the most popular visualization tools, which rely on physical m
etaphors. Graph layout aims to produce aesthetically appealing outputs, with many subjective aims to quantify -- such as minimal overlaps between not related parts (e.g. minimal edge crossings in $d=2$) --, while preserving the symmetries of the network. Altogether, the field of graph drawing became a meeting point of art, physics and computer science \cite{GD}. 

Since the known approaches for the above problems generally lead to computationally expensive NP-hard problems \cite{NP}, the practical implementations were necessarily restricted to advanced approximative heuristic algorithms. Moreover, there was no successful attempt to incorporate network visualization, data ordering and coarse-graining into a common theoretical framework. Since information theory provides ideal tools to quantify the hidden structure in probabilistic data \cite{kinney}, its application to complex networks \cite{slonim,rosvall,map,zanin,bar-yam} is a highly promising field. In this paper, our primary goal is to show an elegant, information theoretic representation theory for the unified solution of network visualization, data ordering and coarse-graining, when the input data has a probabilistic interpretation. 

Usually, in graph theory, the complex system is at the level of abstraction, where each node is a dimensionless object, connected by lines representing their relations, given by the input data. Instead, we study the case in which both the input matrix and the approximative representation is given in the form of probability distributions. This is the routinely considered case of edge weights reflecting the existence, frequency or strength of the interaction, such as in social and technological networks of communication, collaboration and traveling or in biological networks of interacting molecules or species. In our general framework the best representation is selected by the criteria, that it is the hardest to distinguish from the input data. In information theory this is readily obtained by minimizing the relative entropy -- also known as the Kullback-Leibler divergence \cite{KL} -- as a quality function. In the following we show, that the visualization, ordering and coarse-graining of networks are intimately related to each other, being organic parts of a straightforward, unified representation theory. 

\section*{Results}
\subsection*{General network representation theory}
\label{sec:info}
We consider the case, in which the input network is given by the symmetric, node-node co-occurrence $A$ (adjacency) matrix having probabilistic entries $a_{ij}\geq0$. If we start with the $H$ edge-node co-occurrence (incidence) matrix instead, capable to describe hypergraphs as well, then $A\sim H^TH$ is simply given by the elements, $a_{ij}=\frac{1}{h_{**}}\sum_kh_{ki}h_{kj}$. Here and in the following sections asterisks indicate indices, for which the summation was done. Throughout the paper we use the most general form of the input matrices, without assuming their normalization. Similarly, there is no need to normalize the information theoretic measures over $\mathcal{A}$, such as the $S$ information content or the $I$ mutual information given in the Methods section.

The network is represented by a $B$ co-occurrence matrix and a natural way to quantify the quality of the representation is to use the relative entropy. The relative entropy, $D$, measures the extra description length, when $B$ is used to encode the data described by the original matrix, $A$, expressed by
\begin{equation}
D(A||B)=\sum_{ij}{a_{ij}\ln{\frac{a_{ij}b_{**}}{b_{ij}a_{**}}}}\geq0\;.
\label{nn}
\end{equation}   
Although $D(A||B)$ is not a metric and not symmetric in $A$ and $B$, it is an appropriate and widely applied measure of statistical remoteness \cite{mdl}, quantifying the distinguishability of $B$ form $A$. Thus, the highest quality representation is achieved, when the relative entropy approaches $0$, and our general goal is to obtain a $B^*$ representation satisfying
\begin{equation}
B^*={\mathrm{argmin}}_{B}{D(A||B)}\;.
\end{equation}
Since $D(A||B)=H(A,B)-S(A)$, where $H(A,B)$ is the (unnormalized) cross-entropy, we could equivalently minimize the cross-entropy for $B$. Although the minimization of $D(A||B)$ appears in the \emph{minimal discrimination information} approach -- also known as the \emph{minimum cross-entropy} (MinxEnt) approach by Kullback \cite{kullback} --, there the goal is the opposite of ours, namely to find the optimal 'real' distribution, $A$, while the 'approximate' distribution, $B$, is kept fixed. In this sense, our optimization is an \emph{inverse} MinxEnt problem \cite{inv_minxent}. This kind of optimization appears also as a refinement step to improve the importance sampling in Monte Carlo methods (for highly restricted $A$-s), under the name of \emph{cross-entropy method} \cite{rubinstein}. In order to avoid confusion and emphasize the differences, we only use the term of relative entropy in the following.

Although $D(A||B)\geq0$ can be arbitrarily large, there is always available a \emph{trivial} representation by the uncorrelated, \emph{product state}, $B^0$ matrix given by the elements $b^0_{ij}=\frac{1}{a_{**}}a_{i*}a_{*j}$. This way $D_0\equiv D(A||B^0)=I\leq S$ is the $I$ mutual information, thus the optimized value of $D$ can be always normalized with $I$, or alternatively as 
\begin{equation}
\label{eta}
\eta\equiv D/S\leq1\;.
\end{equation}
Here, $\eta$ gives the ratio of the needed extra description length to the optimal description length of the system. In the following applications we use $\eta$ to compare the optimality of the found representations. The optimization of relative entropy is local in the sense, that the global optimum of a network comprising independent subnetworks is also locally optimal for each subnetwork.

The finiteness of $D_0$ ensures, that if $i$ and $j$ are connected in the original network ($a_{ij}>0$), then they are guaranteed to be connected in a meaningful representation as well, enforcing $b_{ij}>0$, since otherwise $D$ would diverge. In the opposite case, when we have a connection in the representation, without a corresponding edge in the original graph ($b_{ij}>0$ while $a_{ij}=0$), $b_{ij}$ does not appear directly in $D$, only globally, through the $b_{**}$ normalization. Nevertheless, the $B$ matrix of the optimal representation (where $D$ is small) is \emph{close} to $A$, since due to Pinsker's inequality the total variation of the normalized distributions is bounded by $D$ \cite{cover}
\begin{equation}
D(A||B) \geq\frac{a_{**}}{2\ln2}\sum_{i,j}{\left(\frac{a_{ij}}{a_{**}}-\frac{b_{ij}}{b_{**}}\right)^2}\;.
\end{equation}
Thus, in the optimal representation of the network all the connected network elements are connected, while having only a strongly suppressed amount of false positive connections. 

\subsection*{Network visualization and data ordering}

Since force-directed layout schemes \cite{force} have an energy or quality function, optimized by efficient techniques borrowed from many-body physics \cite{BH} and computer science \cite{stress_drawing}, graph layout could be in principle serve as a quantitative tool. However, these approaches inherently struggle with an information shortage problem, since the edge weights only provide half the needed data to initialize these techniques. For instance, for the initialization of the widely applied Fruchterman-Reingold \cite{FR} (or for the Kamada-Kawai \cite{KK}) method we need to set both the strength of an attractive force (optimal distance) and a repulsive force (spring constant) between the nodes in order to have a balanced system. Due to the lack of sufficient information, such graph layout techniques become somewhat ill-defined and additional subjective considerations are needed to double the information encoded in the input data, traditionally by a nonlinear transformation of the attractive force parameters onto the parameters of the repulsive force \cite{FR} .  

While in usual graph layout schemes graph nodes are represented by points (without spatial extension) in a $d$-dimensional background space -- connected by (straight, curved or more elaborated) lines --, in our approach network nodes are extended objects, namely probability distributions ($\rho$) over the background space. Edges are defined as overlaps of node distributions. Importantly, in our representation the shape of nodes encodes just that additional set of information, which has been lost and then arbitrarily re-introduced in the above mentioned network visualization methods. In the following we consider the simple case of Gaussian distributions -- having a width of $\sigma$, and norm $h=\int\rho$, see equation (\ref{gauss}) of the Methods section --, but we have also tested the non-differentiable case of a homogeneous distribution in a spherical region of radius $\sigma$. For a given graphical representation the $B$ co-occurrence matrix is built up from the overlaps of the distributions $\rho_i$ and $\rho_j$ -- analogously to the construction of $A$ from $H$ -- as $b_{ij}=\frac{1}{R}\int{\mathrm{d}x^d \rho_i(x)\rho_j(x)}$, where $R=\sum_k\int\mathrm{d}x^d \rho_k(x)$ is an (irrelevant) normalization factor. For a schematic illustration see Fig.~1.

\begin{figure}[!ht]
\vspace*{.05in}
\begin{center}
\includegraphics[width=3.5in,angle=0]{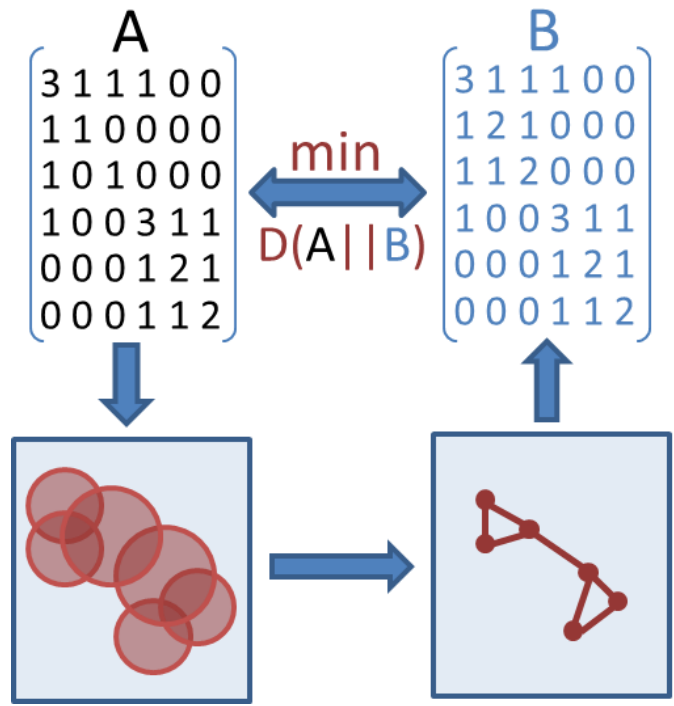}
\end{center}
\caption{
\label{fig_1}
\textbf{Illustration of our network visualization technique.}
For a given $A$ adjacency matrix, we assign a distribution function to each node, from which edge weights ($B$) are calculated based on the overlaps of the distributions. The best layout is given by the representation, which minimizes the $D(A||B)$ description length.
}
\end{figure}

The trivial data representation of $B^0$ can be obtained by an initialization, where all the nodes are at the same position, with the same distribution function (apart from a varying $h_i\propto a_{i*}$ normalization to ensure the proper statistical weight of the rows). This way, initially $D_0=I$ is the mutual information of the input matrix, irrespectively from the chosen distribution function. The numerical optimization can be, in general, straightforwardly carried out by a usual simulated annealing scheme starting with an initialization of $B^0$. Alternatively, in the differentiable case we can also use a Newton-Raphson iteration as in the Kamada-Kawai method \cite{KK} (for details see the Supplementary Information). In terms of the layout, the finiteness of $D_0$ ensures that the connected nodes overlap in the layout as well, even for distributions having a finite support. Moreover, independent parts of the network (nodes or sets of nodes without connections between them) tend to be apart from each other in the layout. Additionally, if two rows (and columns) of the input matrix are proportional to each other, then it is optimal to represent them with the same distribution function in the layout, as though the two rows were fused and moved together.

In the differentiable case, e.g. with Gaussian distributions, our visualization method can be conveniently interpreted as a force-directed method. If the normalized overlap, $b_{ij}/b_{**}$, is smaller at a given edge than the normalized edge weight, $a_{ij}/a_{**}$, then it leads to an attractive force, while the opposite case induces a repulsive force. Fro details see the Supplementary Information. For Gaussian distributions all the nodes overlap in the representations, leading typically to $D>0$ in the optimal representation. However, for distributions with a  finite support, such as the above mentioned homogeneous spheres, perfect layouts with $D=0$ can be easily achieved even for sparse graphs. In $d=2$ dimensions this concept is reminiscent to the celebrated concept of planarity \cite{planarity}. However, our concept can be applied in any dimensions. Furthermore, it goes much beyond planarity, since any network of $I=0$ (e.g. a fully connected graph) is perfectly represented in any dimensions by $B^0$, that is by simply putting all the nodes at the same position. Here we note, that the concept of cross-entropy have already appeared in the field of graph drawing, such as in the methods of refs. \cite{yamada,estevez}. Besides others, the most important difference between these methods and ours is, that in our case the relative entropy (or cross entropy) is calculated over $N\times N$-point distributions for $N$ nodes, while in the related papers \cite{yamada,estevez} only $2$-point distributions of the form $\{a_{ij},1-a_{ij}\}$ were considered.
\begin{figure}[!ht]
\vspace*{.05in}
\begin{center}
\includegraphics[width=6in,angle=0]{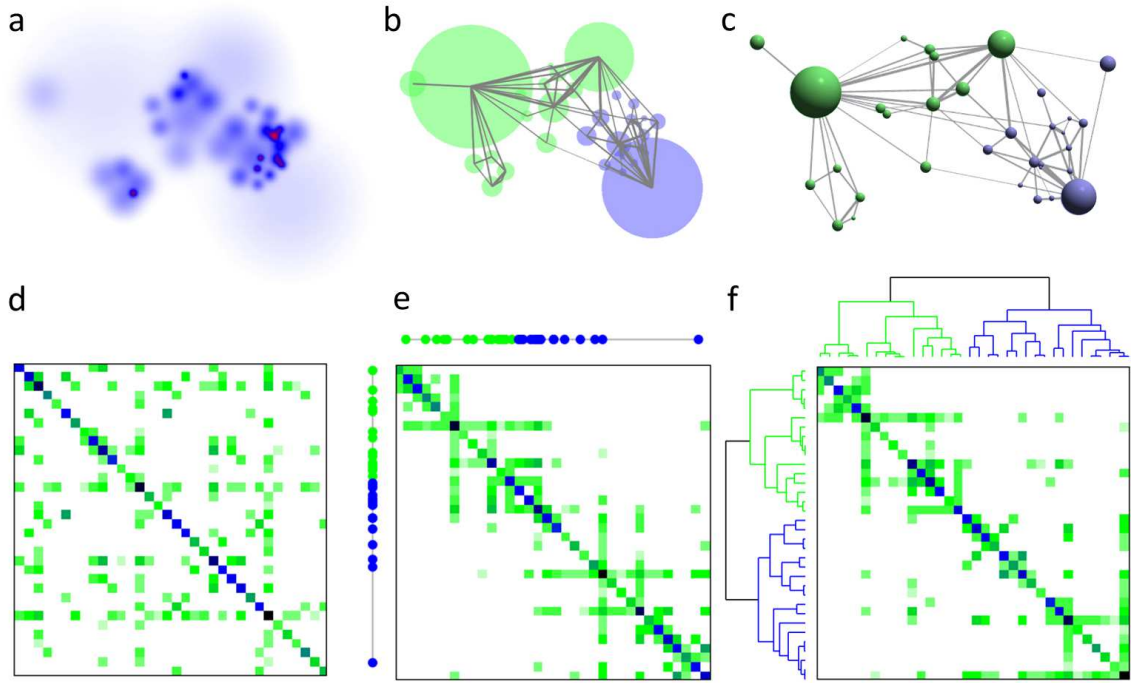}
\end{center}
\caption{
\label{fig_2}
\textbf{Illustration of the power of our unified representation theory on the Zachary karate club network \cite{zachary}.}
The optimal layout ($\eta=2.1\%$, see equation (\ref{eta})) in terms of $d=2$ dimensional Gaussians is shown by a density plot in {\bfseries a} and by circles of radiuses $\sigma_i$ in {\bfseries b}. {\bfseries c} the best layout is obtained in $d=3$ ($\eta=1.7\%$), where the radiuses of the spheres are chosen to be proportional to $\sigma_i$. {\bfseries d} the original data matrix of the network with an arbitrary ordering. {\bfseries e} the $d=1$ layout ($\eta=4.5\%$) yields an optimal ordering of the original data matrix of the network. {\bfseries f} the optimal coarse-gaining of the data matrix yields a tool to zoom out from the network in accordance with the underlying community structure. We note, that the coarse-graining itself does not yield a unique ordering of the nodes, therefore an arbitrary chosen compatible ordering is shown in this panel.
}
\end{figure}

Our method is illustrated in Fig.~2. on the Zachary karate club network \cite{zachary}, which became a cornerstone of graph algorithm testing. It is a weighted social network of friendships between $N_0=34$ members of a karate club at a US university, which fell apart after a debate into two communities indicated by different colors in Fig.~2. While usually the size of the nodes can be chosen arbitrarily, e.g. to illustrate their degree or other characteristics, here the size of the nodes is part of the visualization optimization by reflecting the width of the distribution, indicating relevant information about the layout itself. In fact, the size of a node represents the uncertainty of its position, rather than the importance of the node.    

Our network layout technique works in any dimensions, as illustrated in $d=1,2$ and $3$ in Fig.~2. In each case the communities are clearly recovered and, as expected, the quality of layout becomes better as the dimensionality of the embedding space increases. Nevertheless, the one dimensional case deserves special attention, since it serves as an ordering of the elements as well (after resolving possible degenerations with small perturbations), as illustrated in Fig.~1.e. Although our network layout works only for symmetric co-occurrence matrices, the ordering can be extended for hypergraphs with asymmetric $H$ matrices as well, since the orderings of the two adjacency matrices $HH^T$ and $H^TH$ readily yield an ordering for the rows and columns of the matrix, $H$.

Remarkably, the visualization and ordering is perfectly robust against noise in the input matrix elements. This means, that even if the input $A$ matrix is just the average of a matrix ensemble, where the $a_{ij}$ elements have an (arbitrarily) broad distribution, the optimal representation is the same as it were by optimizing for the whole ensemble simultaneously. This extreme robustness follows straightforwardly from the linearity of the $H(A,B)$ cross-entropy in the $a_{ij}$ matrix elements. Note, however, that the optimal value of the $D(A||B*)$ distinguishability is modified by the noise.

When applying a local scheme (e.g. simulated annealing) for the optimization of the representations, we generally run into computationally hard situations. These correspond to local minima, in which the layout can not be improved by single node updates, since whole parts of the network should be updated (rescaled, rotated or moved over each other), instead. Being a general difficulty in many optimization problems, it was expected to be insurmountable also in our approach. In the following we show, that the relative entropy based coarse-graining scheme -- given in the next section -- can, in practice, efficiently help us trough these difficulties in polynomial time.

\subsection*{Coarse-graining of networks}

Since it is generally expected to be an NP-hard problem to find the optimal simplified, coarse-grained description of a network at a given scale, we have to rely on approximate heuristics having a reasonable run-time. In the following we use a local coarse-graining approach, where in each step a pair of rows (and columns) is replaced by coarse-grained rows, giving the best approximative new network $G'$ in terms of $D(G||G')$, where $G$ is the $H$ or $A$ matrix of the initial network. Although the applied formulas and coarse-graining steps are different, the idea of pairwise, information theoretic coarse-graining appeared recently also in the method of ref. \cite{zanin} to detect the presence of meso-scale structures in complex networks.
In our method, for $G=H$ the coarse-graining step means, that instead of the original $k$ and $l$ rows, we use two new rows, being proportional to each other, while the $h_{k*}'=h_{k*}$ and $h_{l*}'=h_{l*}$ probabilities are kept fixed 
\begin{equation}
h_{ki}'=h_{k*}\frac{h_{ki}+h_{li}}{h_{k*}+h_{l*}}\;,\quad 
h_{li}'=h_{l*}\frac{h_{ki}+h_{li}}{h_{k*}+h_{l*}}\;.
\end{equation}
This way the rows involved are replaced by their normalized average. Figuratively, this way 'bonds' are formed between the nodes in each step with a given $D(G||G')$ 'scale'. As in the case of graph layout, if two rows (or columns) are proportional to each other, they can be fused together, since their coarse-graining leads to $D=0$. Consequently, we can alternatively think of the coarse-graining step as a fusion or grouping of the rows involved. For an illustration of the fused data matrices see the lower panels of Fig.~3.a-d. We note, that proportional rows (or columns) can be generally fused together also initially, in the input data, as a prefiltering, before starting to find an optimal representation.

When $G=A$, the coarse-graining step is carried out simultaneously and identically for the rows and columns. The optimal coarse-graining is illustrated in Fig.~2.f for the Zachary karate club network. The heights in the dendrogram indicate the $D$ values of the representations when the fusion step happens. 

\begin{figure}[!ht]
\vspace*{.05in}
\begin{center}
\includegraphics[width=6in,angle=0]{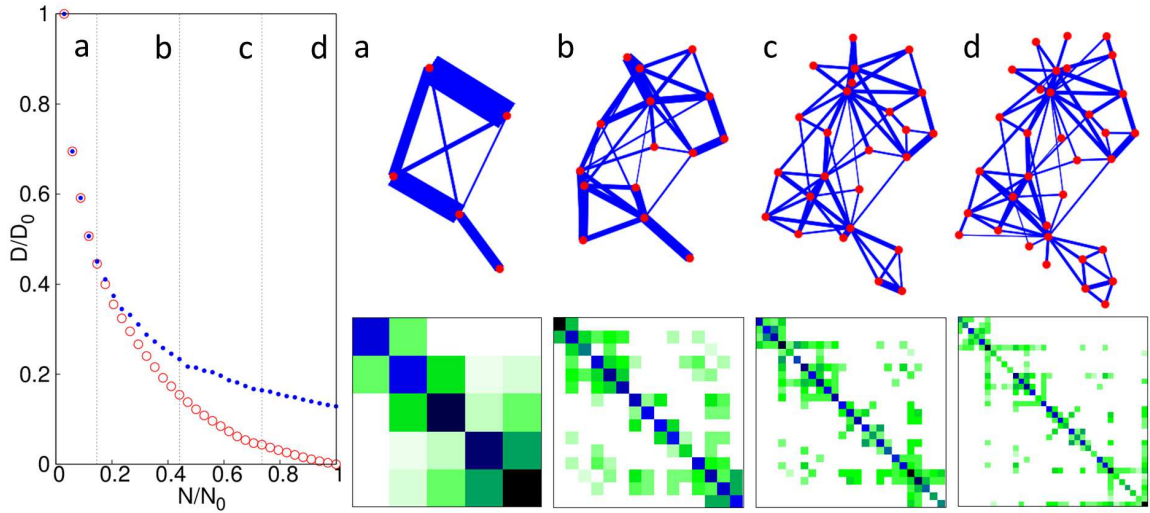}
\end{center}
\caption{
\label{fig_3}
\textbf{Illustration of our hierarchical visualization technique on the Zachary karate club network \cite{zachary}.} In our hierarchical visualization technique the coarse-graining procedure guides the optimization for the layout in a top-down way. As the $N$ number of nodes increases, the relative entropy of both the coarse-grained description (red, $\circ$) and the layout (blue, $\bullet$) decreases. The panels {\bfseries a}-{\bfseries d} show snapshots of the optimal layout and the corresponding coarse-grained input matrix at the level of $N=5, 15, 25$ and $34$ nodes, respectively. For simplicity, here the $h_i$ normalization of each distribution is kept fixed to be $\propto a_{i*}$ during the process, leading finally to $\eta=4.4\%$.
}
\end{figure}

For an alternative formulation of our coarse-graining approach and details on the numerical optimization, see the Supplementary Information. Here we only mention, that the optimization can be generally carried out in roughly $\mathcal{O}(N^3)$ time for $N$ nodes, and it has the following interesting interpretation. For coarse-graining,  $D(A||B)$ is nothing but the amount of lost mutual information between the rows and columns of the input matrix.  In other words, $D(A||B)$ is the amount of lost structural signal during coarse-graining. As a consequence, finally we arrive at $D(A||B)=D_0$ in strong contrast to ref. \cite{zanin}. Prevailingly, this coincides with the above proposed initialization step of our network layout approach. 

\subsection*{Hierarchical layout}

Although the introduced coarse-graining scheme may be of significant interest whenever probabilistic matrices appear, here we focus on its application for network layout, to obtain a hierarchical visualization \cite{ggk, harel_koren, walshaw, yifanhu, bioinfo, grouping}. Our bottom-up coarse-graining results can be readily incorporated into the network layout scheme in a top-down way by initially starting with one node (comprising the whole system), and successively undoing the fusion steps (cutting bonds) until the original system is recovered. Between each such extension step the layout can be optimized as usually. 

We have found, that this hierarchical layout scheme produces significantly improved layouts compared to a local optimization, such as a simple simulated annealing or Newton-Raphson iteration. By incorporating the coarse-graining in a top-down approach, we first arrange the position of the large-scale parts of the network, and refine the picture in later steps only. The refinement steps happen, when the position and extension of the large-scale parts have already been sufficiently optimized. After such a refinement step, the nodes -- moved together so far -- are treated separately. At a given scale (having $N\leq N_0$ nodes), the $D$ value of the coarse-graining provides a lower bound for the $D$ value of the obtainable layout. Our hierarchical visualization approach is illustrated in Fig.~3. with snapshots of the layout and the coarse-grained representation matrices of the Zachary karate club network \cite{zachary} at $N=5, 15, 25$ and $34$. As an illustration on a larger and more challenging network, in Fig.~4. we show the result of the hierarchical visualization on the giant component of the weighted human diseasome network \cite{diseasome}. In this network we have $N_0=516$ nodes, representing diseases, connected by mutually associated genes. The colors indicate the known disease groups, which are found to be colocalized well in the visualization. 
\begin{figure*}[!ht]
\vspace*{.05in}
\begin{center}
\includegraphics[width=7in,angle=0]{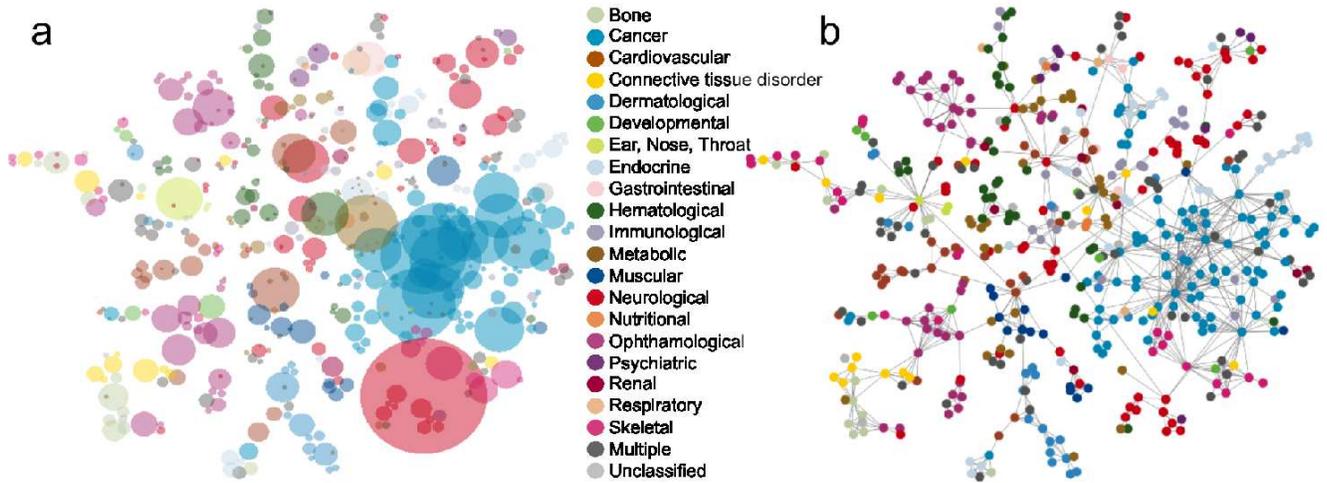}
\end{center}
\caption{
\label{fig_4}
\textbf{Visualization of the human diseasome.} The obtained best layout ($\eta=3.1\%$) by our hierarchical visualization technique of the human diseasome is shown by circles of radiuses $\sigma_i$ in {\bfseries a} and by a traditional graph in {\bfseries b}. The nodes represent diseases, colored according to known disease categories \cite{diseasome}. In the numerical optimization for this network we primarily focused on the positioning of the nodes, thus the optimization for the widths and normalizations was only turned on as a fine-tuning after an initial layout was obtained.
}
\end{figure*}


\section*{Discussion}
\label{sec:discussion}

In this paper, we have introduced a unified, information theoretic solution for the long-standing problems of matrix ordering, network visualization and data coarse-graining. In the presented applications, the steps of the applied algorithms were derived in an \emph{ab inito} way from the first principles of the our framework, in strong contrast to the large variety of existing algorithms, lacking such an underlying theory. First, we demonstrated that the minimization of relative information yields a novel visualization technique, while representing the $A$ input matrix by the $B$ co-occurrence matrix of extended distributions, embedded in a $d$-dimensional space. As another application of the same approach, we obtained a hierarchical coarse-graining scheme, when the input matrix is represented by its subsequently coarse-grained versions. Since these applications are two sides of the same representation theory, they turned out to be superbly compatible, leading to an even more powerful hierarchical visualization technique, illustrated on the real-world example of the human diseasome network. Although we have focused on the visualization in $d$-dimensional continuous space, the representation theory can be applied more generally, incorporating also the case of discrete embedding spaces. As a possible future application, the optimal embedding of a (sub)graph into another graph. 

We have shown that our relative entropy-based visualization with e.g. Gaussian node distributions can be naturally interpreted as a force-directed method. Traditional force directed methods prompted huge efforts on the computational side to achieve scalable algorithms applicable for the large data sets in real life. Here we naturally can not and do not wish to compete with such advanced techniques, but we believe that our approach can be a good starting point for further scalable implementations. We have also demonstrated, that network visualization is already interesting in one dimension yielding an ordering for the elements of the network. Our efficient coarse-graining scheme can also serve as an unbiased, resolution-limit-free, starting point for the infamously challenging problem of community detection by selecting the best cut of the dendrogram based on an appropriately chosen criteria. 

Our data representation framework has a broad applicability, starting form either the node-node or edge-edge adjacency matrices or the edge-node incidence matrix of weighted networks, incorporating also the cases of bipartite graphs and hypergraphs. By applying our unified network representation theory we hope to get closer to discover knowledge from the huge data matrices in science, also in complicated cases beyond the limits of existing heuristic algorithms. Since in this paper our primary intention was merely to demonstrate a proof of concept study of our theoretical framework, the more detailed analysis of interesting complex networks will be the subject of forthcoming articles.

\section*{Methods}
\footnotesize
For details of the numerical optimization for visualization and coarse-graining see the Supplementary Information. Here we briefly overview the most important notations used. The (unnormalized) Shannon entropy of $\mathcal{A}$, expressing the amount of information in the system is given by $S=-\sum_{ij}a_{ij}\ln{\frac{a_{ij}}{a_{**}}}$, while the $I$ mutual information between the rows and columns of $A$ is given by $I=\sum_{ij}a_{ij}\ln{\frac{a_{ij}a_{**}}{a_{i*}^2}}$. 
The parametrization of the Gaussian distributions used in the layout is the following in $d$-dimensions
\begin{equation}
\label{gauss}
\rho(\{x^0_i\},\sigma,n)=\frac{n}{\sqrt{(2\pi)^d}}\exp{\left(-\frac{\sum_{i=1}^d(x_i-x^0_i)^2}{2\sigma^2}\right)}\;.
\end{equation}

\section*{Acknowledgements}
\footnotesize
We are grateful to the members of the LINK-group (www.linkgroup.hu) for useful discussions. This work was supported by the Hungarian National Research Fund under grant Nos. OTKA K109577 and K83314. The research of IAK was supported by the European Union and the State of Hungary, co-financed by the European Social Fund in the framework of T\'AMOP 4.2.4. A/2-11-1-2012-0001 'National Excellence Program'. 

\section*{Author contributions}
\footnotesize IAK and RM conceived the research and ran the numerical simulations. IAK devised and implemented the applied algorithms. IAK and PCs wrote the main manuscript text. All authors reviewed the manuscript.

\section*{Additional information}
\footnotesize
Competing financial interests
The authors declare no competing financial interests.
Correspondence should be addressed to IAK.

{\bf Supplementary Information} accompanies this paper at {\texttt{http://www.nature.com/naturecommunications}}


\clearpage

\section{Supplementary Information}

\section{Numerical optimization for visualization} For the numerical optimization of the network layout, we have implemented a simple, general purpose simulated annealing scheme. For Gaussian distributions we have also worked out a much faster Newton-Raphson update, which has been also applied in the Kamada-Kawai method. In practice, we used a separate Newton-Raphson iteration step for the $d$ coordinates of the nodes in $d$ spatial dimensions and for the $\sigma_i$ widths and $h_i$ normalizations of the distributions. 

In each iteration step of the Newton-Raphson method, the node with the largest gradient amplitude ($||J||$) was updated in the direction and with a parameter step size, obtained by the second derivative matrix, $\mathcal{F}$ as $-\mathcal{F}^{-1}J$. Since $\mathcal{F}$ is not always positive definite, special care was needed when the relative entropy increased in such a step. In such a case, a sufficiently small step size was applied in the direction of the gradient vector, instead. This way our technique has the same computational complexity as the widely applied Kamada-Kawai method (after the initial calculation of pairwise graph-theoretic distances).

While the original, input matrix is given by $A$, the visualization generates the matrix of the pairwise overlaps of the node distributions, marked as $B$. In our approach we minimize the relative entropy between the two distributions, $D(A||B)$, which measures the extra description length, when $B$ is used to encode the data described by $A$,
\begin{equation}
\label{dab}
D(A||B)=\sum_{ij}{a_{ij}\ln{\frac{a_{ij}b_{**}}{b_{ij}a_{**}}}}\geq0\;.
\end{equation}   
Here an asterisks indicates and index for which we summed up. 
During optimization the $a_{ij}$ matrix elements of the $A$ input matrix were kept fixed, while the values of $b_{ij}$ changed due to the variation of the $x_i,y_i,\sigma_i$ (and $h_i$) parameters of the $d$-dimensional Gaussian distributions of the nodes, given by
\begin{equation}
\label{gauss}
\rho(\{x^0_i\},\sigma,n)=\frac{n}{\sqrt{(2\pi)^d}}\exp{\left(-\frac{\sum_{i=1}^d(x_i-x^0_i)^2}{2\sigma^2}\right)}\;.
\end{equation}
The overlap matrix elements of the node distributions were given by
\begin{equation}
b_{ij}=\frac{h_ih_j}{2\pi(\sigma_i^2+\sigma_j^2)}\exp{\left(-\frac{(x_i-x_j)^2+(y_i-y_j)^2}{2(\sigma_i^2+\sigma_j^2)}\right)}\;.\\
\end{equation}

\subsection{Details of the Newton-Raphson update in 2 dimensions}
For a Newton-Raphson iteration step we need to calculate the first and second derivatives of the $D$ relative entropy as the function of the parameters of each node. 

\subsubsection{Updating the coordinates}
When differentiating according to the coordinates, we obtain
\begin{eqnarray}
\frac{\partial b_{kj}}{\partial x_k}&=&-\frac{x_k-x_j}{\sigma_k^2+\sigma_j^2}b_{kj}\;,\\
\frac{\partial D}{\partial x_k}&=&-2a_{**}\sum_j\frac{x_k-x_j}{\sigma_k^2+\sigma_j^2}\left(\frac{b_{kj}}{b_{**}}-\frac{a_{kj}}{a_{**}}\right)\;.
\end{eqnarray}
From this we can see, that $\frac{b_{kj}}{b_{**}}>\frac{a_{kj}}{a_{**}}$ induces a repulsive force, while the opposite case leads to an attractive force.
In order to have an efficient numerical implementation we introduce the following variables.
\begin{eqnarray}
\alpha^x_{kj}&=&-2a_{**}\frac{x_k-x_j}{\sigma_k^2+\sigma_j^2}b_{kj}\;,\\
\beta^x_{kj}&=&-2\frac{x_k-x_j}{\sigma_k^2+\sigma_j^2}a_{kj}\;.
\end{eqnarray}
Here the superscript $x$ indicates, that we now consider the $x$ direction in the formulas. This way the $J$ gradient vector has the following $x$-component
\begin{equation}
j_x\equiv\frac{\partial D}{\partial x_k}=\frac{\alpha^x_{k*}}{b_{**}}-\beta^x_{k*}\;.
\end{equation}
Consequently, while using $\alpha^x$ and $\beta^x$, $\frac{\partial D}{\partial x_k}$ can be calculated in $\mathcal{O}(N)$ time $\forall k$, instead of the $\mathcal{O}(N^2)$ approach of a direct evaluation. For the $y$ direction the same formulas apply with the substitution, $x\leftrightarrow y$.

During the Newton-Raphson method that node, $k$, was updated, for which $||J||\equiv j_x^2+j_y^2$ was the largest.
The $\mathcal{F}$ second derivative matrix had the following elements at a given node, $k$.
\begin{equation}
\begin{split}
f_{xx}\equiv\frac{\partial^2 D}{\partial x_k^2}=-2\frac{a_{**}}{b_{**}}\sum_j\frac{b_{kj}}{\sigma_k^2+\sigma_j^2}+ 2\sum_j\frac{a_{kj}}{\sigma_k^2+\sigma_j^2}\\ -\frac{(\alpha^x_{k*})^2}{2a_{**}b_{**}^2} -\frac{1}{b_{**}}\sum_j\frac{x_k-x_j}{\sigma_k^2+\sigma_j^2}\alpha^x_{kj}\;.
\end{split}
\end{equation}
\begin{equation}
\begin{split}
f_{xy}\equiv\frac{\partial^2 D}{\partial x_k\partial y_k}=-2\frac{\alpha^x_{k*}\alpha^y_{k*}}{2a_{**}b_{**}^2}-\frac{1}{b_{**}}\sum_j\frac{y_k-y_j}{\sigma_k^2+\sigma_j^2}\alpha^x_{kj}\;.
\end{split}
\end{equation}
In the Newton-Raphson method the step size in the $x$ and $y$ directions were automatically given by the vector $-\mathcal{F}^{-1}J$, if $\Delta\equiv f_{xx}f_{yy}-f_{xy}^2\neq0$. As a result, in the $x$- and $y$-directions we obtained
\begin{equation}
\delta_x=\frac{1}{\Delta}(f_{xy}j_y-f_{yy}j_x),\; \delta_y=\frac{1}{\Delta}(f_{xy}j_x-f_{xx}j_y)\;.
\end{equation}
Since $\mathcal{F}$ is not always positive definite (not even in the traditional Kamada-Kawai method), special care was needed when the relative entropy increased in such a step. In such a case, a sufficiently small step size was applied in the direction of the gradient vector, instead of the direction given by $\mathcal{F}$. In our implementation we started with the same step size as before and iteratively kept dividing it by two, until the relative entropy decreased.

\subsubsection{Updating the widths}
The widths were updated separately in a similar manner (there was only one variable at each node). In order to have an efficient implementation, we first introduced the following variable
\begin{equation}
\gamma_{kj}=\frac{\sigma_k}{\sigma_k^2+\sigma_j^2}\left(\frac{(x_k-x_j)^2+(y_k-y_j)^2}{\sigma_k^2+\sigma_j^2}-2\right)\;,
\end{equation}
with which
\begin{equation}
\frac{\partial b_{kj}}{\partial \sigma_k}=b_{kj}\gamma_{kj}\;.
\end{equation}
\begin{equation}
\frac{\partial D}{\partial \sigma_k}=-2\sum_ja_{kj}\gamma_{kj}+2\frac{a_{**}}{b_{**}}\sum_jb_{kj}\gamma_{kj}\;.
\end{equation}
The second derivative at a given node $k$ was 
\begin{equation}
\begin{split}
\frac{\partial^2 D}{\partial \sigma_k^2}=-2\sum_j\frac{a_{kj}}{\sigma_k^2+\sigma_j^2}\epsilon_{kj} - 2\frac{a_{**}}{b_{**}}\sum_j\frac{b_{kj}}{\sigma_k^2+\sigma_j^2}\epsilon_{kj}\\-
\frac{4a_{**}}{b_{**}^2}\left(\sum_jb_{kj}\gamma_{kj}\right)^2+\frac{2a_{**}}{b_{**}}\sum_jb_{kj}\gamma_{kj}^2\;,
\end{split}
\end{equation}
where we used the notation
\begin{equation}
\epsilon_{kj}=\gamma_{kj}\frac{\sigma_j^2-2\sigma_k^2}{\sigma_k}-\frac{2\sigma_k^2}{\sigma_k^2+\sigma_j^2}\;.
\end{equation}

\subsubsection{Updating the normalizations}
Although in many applications it is more natural to keep the normalizations fixed at their original value, it generally leads to improved representations if we update the $h_i$ normalization values as well during the optimization, so we provide here the details for these steps. 
\begin{eqnarray}
\frac{\partial b_{kj}}{\partial h_k}&=&\frac{b_{kj}}{h_k}\;,\\
\frac{\partial D}{\partial h_k}&=&\frac{1}{b_{**}}\frac{2a_{**}b_{k*}}{h_k}-\frac{2a_{k*}}{h_k}\;.
\end{eqnarray}
The second derivative at a given node $k$ was
\begin{equation}
\begin{split}
\frac{\partial^2 D}{\partial h_k^2}=-\frac{4a_{**}b_{k*}^2}{h_k^2b_{**}^2}+\frac{2a_{k*}}{h_k^2}\;.
\end{split}
\end{equation}

\subsection{The case of diagonal elements} 
The above formulas hold for the diagonal $b_{ii}$ self-overlap elements as well. However, the $b_{ii}$ values do not change during repositioning the nodes, but only by updating the $\sigma_i$ widths or $h_i$ normalizations of the Gaussians. Nevertheless, in practice, special care may be needed for the diagonal elements, describing the probability of the co-occurrence of an element with itself. If the nodes represent individual entities in $A$, rather than some properties or groups, then such self co-occurrences are impossible leading to $a_{ii}\equiv 0$, which can be included in the representation scheme as well, by requiring $b_{ii}\equiv0$. While the solution of this case is rather straightforward, for the sake of simplicity we omitted its detailed study.

\section{Alternative formulation of coarse-graining}
In this section we show a simple, intuitive interpretation of our coarse-graining approach.
\subsection{Coarse-graining the rows}
A grouping or coarse-graining, $M$, of the rows of the input matrix $A$ can be generally described by the fusion matrix $U$ as 
\begin{equation}
m_{ij}=\sum_ku_{ik}a_{kj}\;,
\end{equation}
where $u_{*k}=1$, $\forall k$. Instead of this reduced size matrix, in our coarse-graining we used a (practically equivalent), averaged out representation, $B$, of the original size given by the elements
\begin{equation}
b_{ij}=a_{i*}\sum_ku_{ki}\frac{m_{kj}}{m_{k*}}\;.
\end{equation}
By considering partitionings of the rows, without overlaps, each row $i$ had a unique label $\sigma(i)$ yielding its cluster. With this notation
\begin{equation}
b_{ij}=a_{i*}\frac{m_{\sigma(i)j}}{m_{\sigma(i)*}}\;.
\end{equation}
By substituting this into Eq. (\ref{dab}), and changing the indices $\sigma(i)\to i$ we arrive at
\begin{equation}
D=\sum_{ij} a_{ij}\ln\frac{a_{ij}}{a_{i*}}-\sum_{ij} m_{ij}\ln\frac{m_{ij}}{m_{i*}}\;,
\end{equation}
which is simply 
\begin{equation}
D=I(R,C)-I(r,C)\;,
\end{equation}
where $r$ means the coarse-grained set of rows, $R$. Since the $I(R,C)\equiv D_0$ mutual information can be interpreted as the amount of structural 'signal' in the original data, $D$ is the amount of lost structural signal during coarse-graining.

\subsection{Coarse-graining both the rows and columns}
The simultaneous coarse-graining of the rows and columns of $A$ was given by the matrix elements
\begin{equation}
w_{ij}=\sum_ku_{ik}v_{jl}a_{kl}\;,
\end{equation}
where $v_{*l}=1$, $\forall l$. The averaged out representation, $B$, of the original size was given in this case by the elements
\begin{equation}
b_{ij}=a_{i*}a_{*j}\sum_{kl}u_{ki}v_{lj}\frac{w_{kl}}{w_{k*}w_{*l}}\;.
\end{equation}
By considering partitionings, this can be written in the form of
\begin{equation}
b_{ij}=a_{i*}a_{*j}\frac{w_{\sigma(k)\sigma(l)}}{w_{\sigma(k)*}w_{*\sigma(l)}}\;.
\end{equation}
By substituting this into Eq. (\ref{dab}), and changing the indices as before, we arrive at
\begin{equation}
D=\sum_{ij} a_{ij}\ln\frac{a_{ij}}{a_{i*}a_{*j}}-\sum_{ij} w_{ij}\ln\frac{w_{ij}}{w_{i*}w_{*j}}\;,
\end{equation}
which is simply 
\begin{equation}
D=I(R,C)-I(r,c)\;,
\end{equation}
where $r$ ($c$) means the coarse-grained $R$ ($C$). Thus, it is true also in this case, that $D$ can be interpreted as the amount of lost structural signal during coarse-graining. For a graphical representation of these considerations see Fig.~1. of the Supplementary Information.

\begin{figure}[!ht]
\vspace*{.05in}
\begin{center}
\includegraphics[width=3.42in,angle=0]{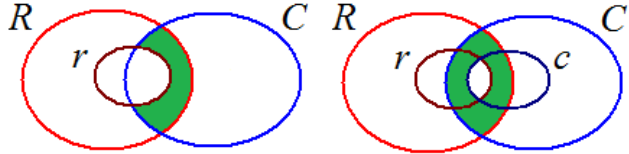}
\end{center}
\caption{
\label{fig_1}
{\bf Graphical representation of the relative entropy, $D$, used in the hierarchical coarse-graining.} In the information content diagram the shaded (green) area indicates $D$, which is found to be the amount of the lost mutual information between the rows ($R$) and the columns  ($C$). The coarse-grained rows and columns are denoted by $r$ and $c$, respectively. Left: coarse-graining for the rows ($R$) only. Right: simultaneous coarse-graining of the rows and columns.}
\end{figure}

\section{Greedy optimization for coarse-graining}

Since in a coarse-graining step the $B$ representation matrix is modified, for the remaining steps the $D$ difference should be updated if at least one member of the pair is neighbor of the fused elements. Although it seems to be somewhat tedious in a later step to measure $D$ always from the original input matrix for a given $(k,l)$ pair, there is a simple rule to calculate this from the actually existing coarse-grained data alone. If $D_k$ and $D_l$ are the values when the rows (and columns) $k$ and $l$ were formed via fusion (being zero initially), then from the apparent $D'$ value -- measuring the formation of a bond directly from the coarse-grained rows $k$ and $l$ -- we got 
\begin{equation}
D=D_k+D_l+D'\;.
\end{equation}
This results is valid for both the coarse-graining of the rows and for the simultaneous coarse-graining of both the rows and columns.

\subsection{Coarse-graining of the rows}
In the following we overview the numerical details of coarse-graining the rows of a matrix with $N_r$ rows and $N_c$ columns. For each pair of rows the $\delta$ difference of the $D$ relative entropy value for the fusion step could be calculated independently from the other pairs. Thus after a fusion step only the $\delta$ values of the new row with the rest of the rows were needed to be calculated in $\mathcal{O}(N_c)$ time. Since in each step the pair with the lowest $\delta$ value was fused, we needed to select the lowest value before each step, which could be conveniently done with a binary heap data structure in $\mathcal{O}(\ln{N_r})$ time. Altogether we finished in $\mathcal{O}\left(N_r^2N_c)\right)$ time.

\subsection{Coarse-graining of both the rows and columns}
In the following we overview the numerical details of the simultaneous coarse-graining of both the rows and columns of a symmetric matrix with $N$ rows and columns. In this case the fusion of two node pairs is generally not independent, thus besides calculating the $\delta$ values of the new row, all the other values may be needed to be updated. Fortunately, this can be done in constant time between rows $i$ and $j$. After the fusion of rows $a$ and $b$, $\delta_{ij}$ must be increased by $2\Delta_{ij}$, where
\begin{equation}
\begin{split}
\Delta_{ij}=-w_{ia}\ln w_{ia} -w_{ja}\ln w_{ja}-w_{ib}\ln w_{ib}-w_{jb}\ln w_{jb} \\
+(w_{ia}+w_{ja})\ln (w_{ia}+w_{ja})+(w_{ib}+w_{jb})\ln (w_{ib}+w_{jb})\\
+(w_{ia}+w_{ib})\ln (w_{ia}+w_{ib})+(w_{ja}+w_{jb})\ln (w_{ja}+w_{jb})\\
-(w_{ia}+w_{ja}+w_{ib}+w_{jb})\ln (w_{ia}+w_{ja}+w_{ib}+w_{jb})\;.
\end{split}
\end{equation}
Altogether the whole process took $\mathcal{O}(N^3)$ time.


\section{Basic notations}
With the notations $m_{ij}=\sum_ku_{ik}a_{kj}$, $n_{ij}=\sum_ka_{ik}v_{jk}$ and $w_{ij}=\sum_{kl}u_{ik}a_{kl}v_{jl}$ the relevant entropy measures can be expressed as follows.
\begin{equation}
S(R)=-\sum_ia_{i*}\ln{\frac{a_{i*}}{a_{**}}},\;
S(C)=-\sum_ia_{*i}\ln{\frac{a_{*i}}{a_{**}}}
\end{equation}
\begin{equation}
S(r)=-\sum_im_{i*}\ln{\frac{m_{i*}}{a_{**}}}\,;
S(c)=-\sum_in_{*i}\ln{\frac{n_{*i}}{a_{**}}}
\end{equation}
\begin{equation}
S(r,C)=-\sum_{ij}m_{ij}\ln{\frac{m_{ij}}{a_{**}}},\;
S(R,c)=-\sum_{ij}n_{ij}\ln{\frac{n_{ij}}{a_{**}}}
\end{equation}
\begin{equation}
S(R,C)=-\sum_{ij}a_{ij}\ln{\frac{a_{ij}}{a_{**}}},\;
S(r,c)=-\sum_{ij}w_{ij}\ln{\frac{w_{ij}}{a_{**}}}
\end{equation}
From these we could deduce the used measures of mutual information for any $X$ and $Y$ as $I(X,Y)=S(X)+S(Y)-S(X,Y)$.

\end{document}